# Reflective Metalenses for Near-Infrared Wavelengths Based on Silicon Nanorods

Iftekhar Ahmed
Department of Electrical and Electronic Engineering
Shahjalal University of Science and Technology
Sylhet - 3114, Bangaldesh
https://orcid.org/0000-0000-0000-0000

K. B. M. Sharif Mahmood
Department of Electrical Engineering
Texas State University
San Marcos, TX - 78666, USA
https://orcid.org/0000-0002-5209-4024

Tanvin Tamanna
Department of Mathematics
Texas State University
San Marcos, TX – 78666, USA
https://orcid.org/0009-0006-9006-5220

*Abstract—* This paper presents the design and analysis of reflective metalenses optimized for a 900 nm wavelength, using silicon nanorods as the primary components. The metalens consists of unit cells, each containing a thin silicon rod. Lumerical FDTD software is used to create a detailed library that connects the dimensions of these nanorods with their respective phase shifts and reflectance properties. The nanorods are placed on a 70 nm thick silicon dioxide ($SiO_2$) layer, with a 50 nm thick gold (Au) reflective layer underneath. In contrast to conventional transmissive metalenses, our design accomplishes complete $2\pi$ phase control via geometry-optimized silicon nanorods, facilitating compact reflective optics. By carefully examining the impact of nanorod dimensions on optical performance, the main goal is to optimize the metalens design for enhanced light focusing and manipulation, which could benefit imaging systems, optical communications, and sensor technology.

## I. Introduction

Optical lenses have been widely used in materials analysis, photolithography, sensing and detection, and high-resolution imaging (1). The resolution of optical lenses is restricted to the order of the wavelength of light used to picture the object due to the diffraction limit of light, which is brought on by the loss of evanescent waves in the far field that contain high spatial frequency information (2). Metalenses, known as metasurface-based flat optics, provide advanced optical functions in a compact, effective alternative to traditional lenses (3,4). A two-dimensional metamaterial called all-dielectric metalens, that consists of numerous dielectric nano-antennas, has advantages that allow for the develoment of minuscule integrated lenses (5). As of right now, a variety of transmissive and reflecting metalenses in the visible, infrared, and terahertz wavelength range have been exhibited (6). Numerous dielectric materials have been used to display various metalenses structures, including silicon (Si) (7), silicon dioxide ($SiO2$) (8), titanium oxide ($TiO2$) (9), and gallium nitride (GaN) (10). Among them silicon dioxide ($SiO2$) is promising material for metalens application because its high reflectivity properties in near infrared spectrm. Gold (Au) or silver (Ag)-based metallic metasurfaces perform well from THz to near-IR (6).

Achieving achromatic focusing with metalenses requires independent phase control at each wavelength, which is challenging when phase shifts are obtained by varying only the geometry or orientation of a single set of meta-atoms(4). This limitation leads to chromatic aberration, prompting the use of various models like nanoslits(11), nanoholes(12,13), and graphene ribbons (14-16) to achieve full 0–2π phase coverage.

While most prior metasurface research has focused on transmissive configurations, reflective metalenses remain underexplored, especially in the near-infrared range using single-layer dielectric architectures. Our work addresses this gap by proposing a fabrication-conscious design using silicon nanorods on a silicon dioxide layer backed by a gold mirror. This structure simplifies fabrication while offering strong optical performance.

In this study, we designed a reflective metalens using silicon nanorods optimized for operation at a near-infrared wavelength of 900 nm. A silicon dioxide ($SiO_2$) layer was employed as the substrate, and a gold (Au) layer was introduced beneath the $SiO_2$ substrate. The goal of this study is to design a reflective metalens consisting of cylindrical nanorods. We obtained the phase profile by varying height and radius parameters via parametic sweep. We compared the simulated phase profile with the target phase profile. The radius and the arrangement of the nanorods are merged to create a desired phase profile on the metalens surface.

## II. Design of Metalens

### A. Methodology

Metalenses are built from arrays of unit cells with subwavelength structures, and these structures act as phase shifters that modify the wavefront of incoming light. Each unit cell is designed with a specific geometry, and its shape determines the amount of phase shift it applies. These phase shifters provide full control over the light phase from 0 to 360 degrees, and this allows the wave to be steered and focused at a desired point. We use nanocylinders as unit cells, and we set their radii based on the target focal length. By adjusting the radius of the nanocylinders, we control the phase and amplitude of the reflected wave.

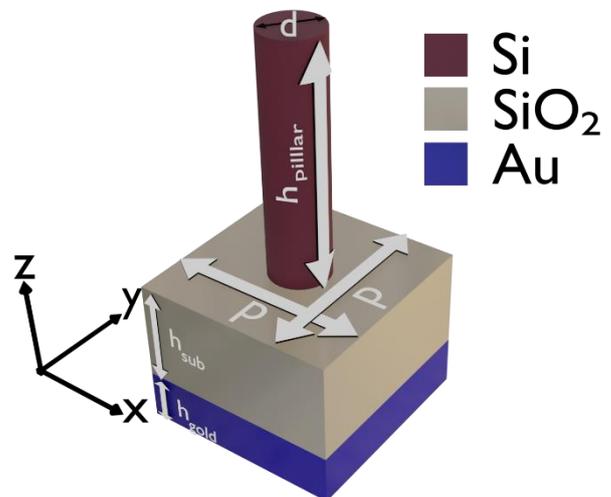

Fig. 1. Unit cell for designing reflective metalens

The phase at each point on the metalens is given by:

$$\phi(x,y) = \frac{2\pi}{\lambda}\left(f - \sqrt{f^2 + x^2 + y^2}\right) \quad (1)$$

Where $x$ and $y$ are the unit cell coordinates, $f$ is the focal length, and $\lambda$ is the wavelength in free space. To focus the beam using a cylindrical lens, we only apply power in the y direction, and the cylindrical lens alone would focus the light into a line along the x-axis.

### B. Unit Cell Design

We used silicon nanorod to design the unit cell because it supports compact structures and provides high phase tunability. The full cell consists of an array of amorphous silicon nanorods and a gold ground plane, because of gold's excellent reflective characteristics in the infrared spectrum. A silicon dioxide ($SiO_2$) spacer was placed between these two layers, primarily due to its compatibility with silicon, which minimizes the likelihood of lattice mismatch. Fig. 1 shows the side view of the unit cell. In this figure, $d$ is the diameter of the nanorod, $h_{pillar}$ is the height of the pillar, $h_{sub}$ is the substrate thickness, $h_{gold}$ is the thickness of the reflective plate and $P_x$ and $P_y$ are the periods in the x and y directions. We set the values as: $d = 160$ nm, $h_{pillar} = 1100$ nm, $h_{sub} = 70$ nm, $h_{gold} = 50$ nm and $(P_x = P_y = 400)$ nm.

We used Lumerical FDTD to simulate the unit cell. A plane wave source illuminated the cell along the z-axis at a wavelength of 900 nm. To measure the performance of the unit cell, we placed a DFT monitor above the source to record the reflected wave and locate the focal point. We analyzed the cross section to assess the reflection and phase response. We also performed a parameter sweep on the radius to build a phase library. This helped us track how reflection and phase changed as we changed the radius. We aimed to cover a $2\pi$ phase shift across the full radius range. We also swept the pillar height to find the value that gave the highest reflection.

### C. Full Lens Design

Once we know the radius distribution, we create the full lens and simulate it in FDTD. We import the phase versus radius data from the unit cell simulation library. This data allows us to do phase-to-radius mapping and place each nanorod at the correct position with the correct radius. Figure 3 shows the design of the reflective metalens, which is made to focus right circularly polarized (RCP) light at the center. The wavelength is set at 900 nm, and the focal length is designed to be 100 μm.

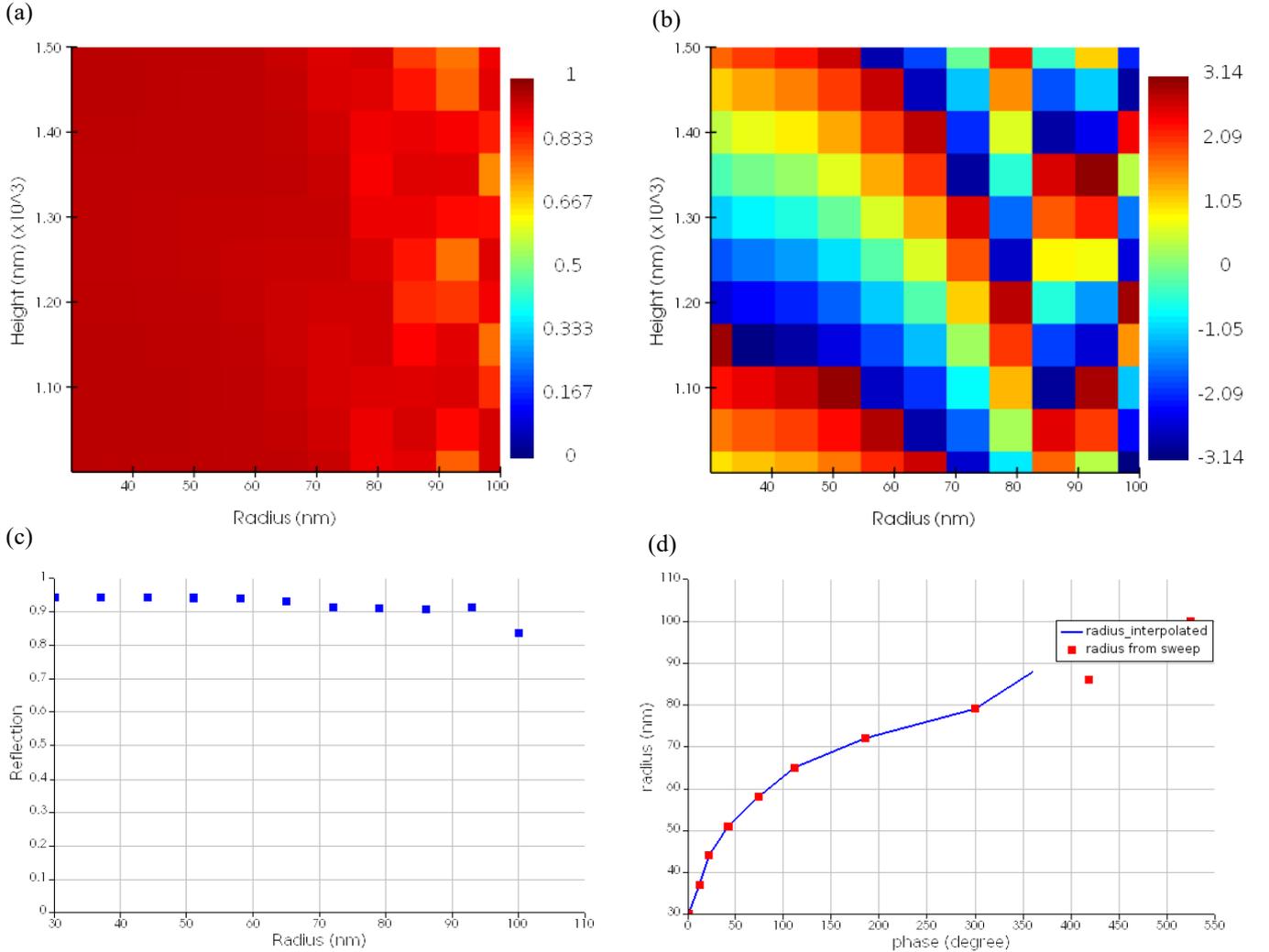

Fig. 2. (a) & (b) 2D maps of phase and reflection based on changes in height and radius of the rods. (c) Reflection for the case when h = 1.1 μm. (d) Required radius to produce a certain phase when h = 1.1 μm.

## III. RESULTS AND DISCUSSION.

### A. Unit Cell Simulation

We created 2D maps of phase and reflection based on changes in height and radius of the rods, as showed in Fig. 2. (a-b). The reflection stayed high for the full range of radius and height values. The phase change covered more than 2 when the rod height was 1.1 µm or more.

Fig. 2. (c) shows the reflection line plot for the case where the height is 1.1 µm. We also created a plot showing the radius needed for each phase value. This plot is the reverse of the phase versus radius plot. We then interpolated the data using finer points for phase and radius. This helped us match the target phase more closely with the actual phase. This is shown in Fig. 2. (d).

### B. Full Lens Simulation

Using Equation 1, we finally translated the spatial phase profile into a spatial radius (of nanorod) distribution for full lens design. Fig. 3.(a) shows the top view of the reflective metalens refractive index profile. Fig. 3.(b) and Fig. 3.(c) show the simulated electric field intensity profiles ($|E|^2$) in the x-y and x-z planes. The flat metalens shows strong focusing with a numerical aperture (NA) of approximately 0.148 and a focal spot defined by FWHMs of 3.07 µm in the x-direction and 2.67 µm in the y-direction. Focusing performance is evaluated based on the amount of incident light concentrated within a circular area with a radius equal to three times the FWHM of the focal spot. These results are part of an ongoing optimization, and improved focusing is expected in the final design.

## IV. COMPARATIVE ANALYSIS

We performed a detailed comparison between our reflective metalens and previously published designs, focusing on key optical performance parameters. Table 1 presents data such as focal length, spot size, and numerical aperture (NA) for various reflective metalenses made with different materials. Yang et al. (2017) [17] developed a reflective metalens using amorphous silicon nanoblocks on a gold mirror and reported a focal length of 20 µm and a spot size of approximately 1.1 µm. Another design used dielectric Bragg reflector layers and achieved a similar spot size with an NA of 0.6. A different study by Guo et al. (2021) [18] proposed a low-NA reflective metalens with a focal length above 2 µm.

Our current simulation shows a focal length of 74.36 µm with a full-width at half-maximum (FWHM) of 3.07 µm along the x-axis and 2.67 µm along the y-axis. The numerical aperture is calculated as 0.148, based on the lens radius and focal length. This spot is slightly elliptical, which is typical for non-circular phase profiles. Although our design does not yet reach diffraction-limited performance, the result is comparable to earlier metalens designs in terms of basic focusing behaviour. These results are from an early stage of optimization. Further improvement in beam quality and focus size is expected as the design is refined. Our structure also avoids multilayer complexity by using a single-layer layout, which supports easier fabrication and integration.

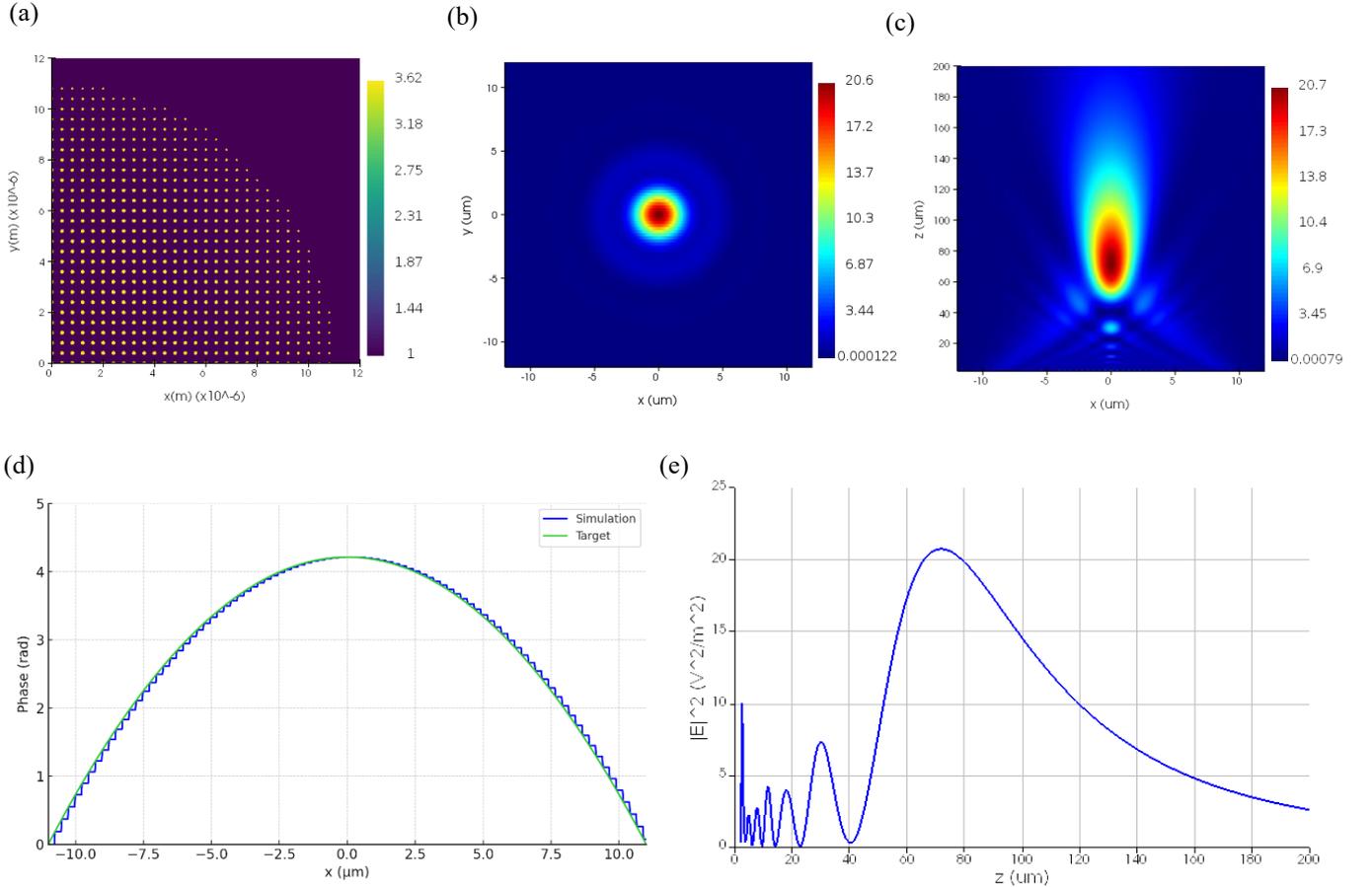

Fig. 3. (a) Refractive Index profile of full lens. (b) & (c) 2D maps of simulated electric field intensity profiles ($|E|^2$) in the x-y and x-z planes. (d) Phase profile (e) Line plot denoting focusing in z direction.

TABLE I. COMPARISON OF REFLECTIVE METALENS DESIGNS

| Study | Wavelength | Focal Length (μm) | FWHM (μm) | NA | Notes |
|---|---|---|---|---|---|
| Yang et al. (2017) | 1550 nm | 20 | ~1.1 | 0.65 | a-Si nanoblocks on gold with spacer |
| Guo et al. (2021) | 633 nm | >2000 | ~2.03 | 0.15 | DBR-based reflective metalens |
| This work (in progress) | 900 nm | 74.36 | 3.07 (x), 2.67 (y) | 0.148 | Initial result; optimization ongoing |

## V. CONCLUSION

We proposed a reflective metalens structure based on dielectric nanorods and simulated its optical performance using the FDTD method. The metalens was designed to achieve a focal length of 74.36 μm at a wavelength of 900 nm by engineering the phase shift through radius variation across the lens surface. The simulated focal spot showed an elliptical shape with FWHMs of 3.07 μm and 2.67 μm in the x and y directions, respectively, and a calculated numerical aperture of 0.148. While the current design is still under optimization, the focusing performance and field distribution confirm effective phase modulation by the nanorod array. Our structure avoids complex multilayer arrangements by using a single-layer layout, which supports simplified fabrication. The findings from this work demonstrate the potential of reflective metalens designs in practical near-infrared optical systems and provide a foundation for further improvement in compact, efficient focusing devices.

## REFERENCES

[1] Progress in the design, nanofabrication, and performance of metalenses, https://doi.org/10.1088/2040-8986/ac44d8
[2] Far-Field Optical Hyperlens Magnifying Sub-Diffraction-Limited Objects Zhaowei Liu,* Hyesog Lee,* Yi Xiong, Cheng Sun, Xiang Zhang
[3] Metalenses: Advances and Applications; DOI: 10.1002/adom.201800554
[4] Reflective metalens with an enhanced off-axis focusing performance; Vol. 30, No. 19/12 Sep 2022/Optics Express 34117; https://doi.org/10.1364/OE.468316
[5] 5.High-efficiency broadband achromatic metalens for near-IR biological imaging window; https://www.nature.com/articles/s41467-021-25797-9
[6] 6. Design of Polarization-Independent Reflective Metalens in the Ultraviolet–Visible Wavelength Region; https://doi.org/10.3390/nano11051243
[7] 7.M.W.Khalid,J.Ha,M.S.E.Hadri,L.Hsu,S.Hemayat,Y.Xiao,A.Sergienko,E.E.Fullerton,andA.Ndao, "Meta-magnetic all-optical helicity dependent switching of ferromagnetic thin films,"Adv.Opt.Mater.12,2301599 (2024)
[8] 8. ,S chen P.Lin, J.Lin,andY.-S.Lin 'Tunableall-dielectricmetalenswithultrahigh-resolutioncharacteristic'
[9] 9. Y.Wang,Q.Chen,W.Yang,Z.Ji,L.Jin,X.Ma,Q.Song,A.Boltasseva,J.Han,V.M.Shalaev,andS.Xiao "High-efficiencybroadbandachromaticmetalensfornear-IRbiologicalimagingwindow,"
[10] 10. B.H.Chen,P.C.Wu,V.-C.Su,Y.-C.Lai,C.H.Chu,I.C.Lee,J.-W.Chen,Y.H.Chen,Y.-C.Lan,C.-H.Kuan,and D.P.Tsai,"GaN metalens for pixel-levelfull-color routing at visible light,"NanoLett.17,6345–6352(2017).
[11] 11. Zhu, Y.; Yuan, W.; Li, W.; Sun, H.; Qi, K.; Yu, Y. TE-polarized design for metallic slit lenses: A way to deep-subwavelength focusing over a broad wavelength range. Opt. Lett. 2018, 43, 206–209.
[12] 12. Zhang, J.; Guo, Z.; Ge, C.; Wang, W.; Li, R.; Sun, Y.; Shen, F.; Qu, S.; Gao, J. Plasmonic focusing lens based on single-turn nano-pinholes array. Opt. Express 2015, 23, 17883–17891
[13] 13. Jia, Y.; Lan, T.; Liu, P.; Li, Z. Polarization-insensitive, high numerical aperture metalens with nanoholes and surface corrugations. Opt. Commun. 2018, 429, 100–105.
[14] 14. Li, Z.; Yao, K.; Xia, F.; Shen, S.; Tian, J.; Liu, Y. Graphene Plasmonic Metasurfaces to Steer Infrared Light. Sci. Rep. 2015, 5, 12423
[15] 15. Zhao,H.;Chen,Z.; Su,F.; Ren, G.; Liu, F.; Yao, J. Terahertz wavefront manipulating by double-layer graphene ribbons metasurface. Opt. Commun. 2017, 402, 523–526.
[16] 16. Ma,W.;Huang, Z.; Bai, X.; Zhang, P.; Liu, Y. Dual-band light focusing using stacked graphene metasurfaces. ACSPhotonics 2017, 4, 1770–1775.
[17] Guo, Q., Zhang, X., Chen, M., & Li, Y. (2021). Ultra-broadband and highly efficient reflective metalens based on multilayer dielectric structures. Nanomaterials, 11(5), 1243. https://doi.org/10.3390/nano11051243
[18] Yang, H., Wang, Y., & Zhang, B. (2017). Sub-diffraction-limited and multifunctional focusing with a reflective metasurface. Scientific Reports, 7(1), 1–9. https://doi.org/10.1038/s41598-017-13004-z